\documentstyle[12pt,epsfig,aaspp]{article}
\lefthead{Gallart}
\righthead{The AGB-bump}

\begin{document}

\title{Observational discovery of the AGB--bump in densely populated color-magnitude diagrams of galaxies and star clusters.}

\author{Carme Gallart}
\affil{The Observatories of the Carnegie Institution of Washington, 813 Santa Barbara St., Pasadena, CA 91101, USA}
\hspace{5.0 truecm} email: carme@ociw.edu

Subject Headings: stars: HR diagram; stars: evolution; stars: AGB and post AGB; galaxies: Local Group; globular clusters: general

\begin{abstract}

The huge photometric databases that are being created for stars in 
the Magellanic Clouds and in Local Group galaxies not only allow 
detailed study of the star formation history of these systems, but 
also provide important information on stellar evolutionary theory. 
Two low-level features above the red-clump in the LMC color-magnitude 
diagram (CMD) have been discussed by Zaritsky \& Lin (1997) and Alcock 
et al. (1997) as the possible signature of an intervening population. 
I conclude that one of the features, which has also been associated 
with the red giant branch bump (RGB--bump) predicted by theory and 
observed in globular clusters, is instead produced during asymtotic 
giant branch (AGB) evolution. I will call it the AGB--bump. In this 
paper, it will be shown that the stellar evolution predictions about 
the position and strength of the AGB--bump are in very good agreement 
with the observed structures in the LMC, M31 and globular clusters. 
The position in the CMD of the other feature is consistent with the 
location of the blue-loops of a few Myr old stars and it has also 
been observed in the CMD of other galaxies with a young stellar 
population.

\end{abstract}

\section{Introduction} \label{intro}

Recently, huge photometric databases are being created for the Magellanic Clouds by the MACHO group (Alcock et al. 1997) and by Zaritsky, Harris \& Thompson (1997). Likewise, an impressive database of Local Group galaxies is being accumulated in the {\it HST} archive. These data routinely show well-known structures in the CMD such as the horizontal-branch (HB) and red-clump (RC), which are key to interpreting the stellar populations. Moreover, other structures which have not been obvious previously, are also apparent in the most populated of these new CMDs. In the case of the LMC, two of these structures, situated above the RC, have been discussed as possible evidence for an intervening stellar population. In this paper, it will be shown that one of these newly identified structures, which I will call the AGB--bump, appears as a consequence of stellar evolution in the AGB phase. Evidence for the other feature being produced by the blue-loops (BL) of a few Myr old stars (Beaulieu \& Sackett 1997: BS) will also be discussed. 

This paper is organized as follows: in Section~\ref{observa} the various features observed in different CMDs near the RC area will be discussed.  In Section~\ref{predice}, the Padova models will be used to show that the AGB--bump is predicted by stellar evolution, in addition to the RGB--bump.  In Section~\ref{compara}, the predictions of the stellar evolutionary models will be compared with observations of the LMC and M31 available in the literature. 
In Section~\ref{glob}, some observations of the AGB--bump in globular clusters will be discussed.

\section{The subtle structures in red giant branch area near the red-clump} \label{observa}

\begin{figure}
\begin{center}
\mbox{\epsfig{file=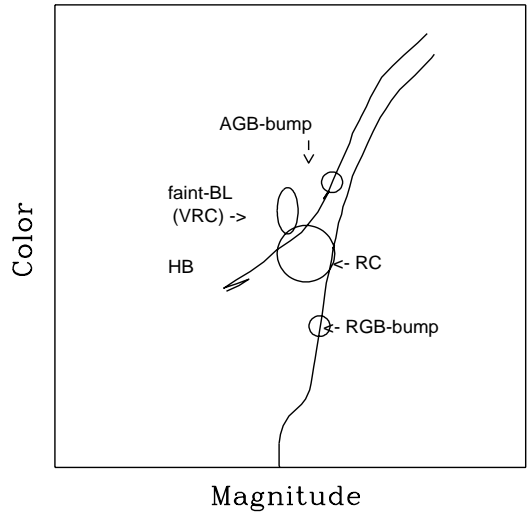,width=8cm,bbllx=3.0truecm,bblly=9.0truecm,bburx=18.0truecm,bbury=18.0truecm}}
\end{center}
\caption[]{Schematic representation of the different structures produced by old and intermediate-age stars near the RC in a composite stellar population.}
\label{esquema}
\end{figure}

 Figure~\ref{esquema} schematically displays the features expected near the RC area of the CMD for a composite stellar population. Three of the structures are well known and thoroughly discussed in the literature: the HB and RC are the locus of the core He-burning phase; the RGB--bump is produced during the ascent of the RGB by low--mass stars (see Section~\ref{predice}). The other two features are the following: 

$\bullet$ AGB--bump: Alcock et al. (1997) noted a ``bump'' superimposed on the RGB and situated above and redwards to the RC in their LMC data. Both ZL and BS identify this structure with the RGB--bump. Nevertheless, for the metallicity of the LMC, the RGB--bump would be observed {\it below} the RC (Fusi Pecci et al. 1990: FP). I will argue that this structure is produced during the AGB evolution and will call it AGB--bump. Other objects that exhibit the AGB--bump in published data include Fornax (Stetson 1997), Andromeda I (Da Costa et al. 1996), M3 (Sandage 1970; Ferraro et al. 1997), the {\it HST} observations of a large number of  globular clusters (Rich et al. 1997) and M31 halo fields. The nature of these stars, rarely noticed as a distinct population in the CMD, has not been discussed so far. Holland, Fahlman \& Richer (1996) noted a ``clump'' above the RC in their M31 {\it HST} CMD, but concluded that the fluctuation of the luminosity function (LF) is not statistically significant.  Note that the structure and position of the AGB--bump will vary depending on the star formation history (SFH) and metallicity of the system. For a single--age population, the AGB--bump is bluer than the RGB. In composite stellar populations it appears to be entangled in the RGB-AGB area.  

$\bullet$ Faint BL -- VRC: the blue--loops produced during the core He-burning phase in a few Myr to 1 Gyr old population (called {\it vertical extension of the RC}, VRC, by Zaritsky \& Lin (1997: ZL). These authors have proposed that this structure, as seen in their LMC CMDs, may be produced by an intervening stellar population toward the LMC. Nevertheless, a similar structure, produced by the faintest BLs of the stars with age $\le$ 1 Gyr, appears in the synthetic CMD shown in Figure~2 of Aparicio et al. (1996), which represents the CMD for a stellar system with constant SFR from 15 Gyr ago to the present time (Fig. 2A) or to 500 Myr ago (Fig. 2B). The same conclusion is reached by BS. Several CMD of galaxies containing a young population also show this feature, including Fornax (Stetson 1997) and Sextans~A (Dohm--Palmer et al. 1997). 

\section{The predictions of stellar evolution for the RGB--bump and the AGB--bump} \label{predice}

\begin{figure}
\begin{center}
\mbox{\epsfig{file=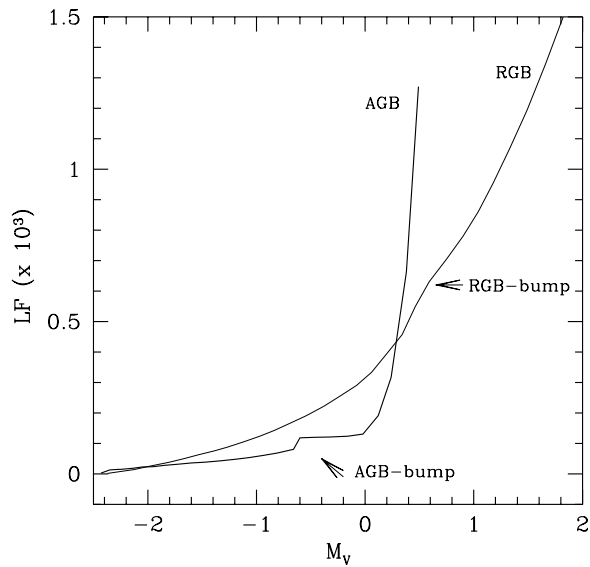,width=8cm,bbllx=1.0truecm,bblly=8.0truecm,bburx=20.0truecm,bbury=20.0truecm}}
\end{center}
\caption[]{Integral luminosity function (LF) for the RGB and AGB, in units of number of stars with magnitude between $M_{V,TRGB}$ and $M_V$ in each evolutionary phase relative to the total number in the population. The positions of the RGB--bump and the AGB--bump are indicated.}
\label{fl004}
\end{figure}

The AGB and the RGB--bump appear as a consequence of analogous processes in the evolution of stars in the AGB and the RGB respectively, when low and intermediate-mass stars approach the Hayashi Line. In both cases,  the expansion and penetration of the convective envelope into the inner layers (when the H- or He-exhausted core contracts and heats up), and its subsequent contraction and recession, slows down the evolution of the star momentarily. This  produces changes in the LF of the stars in the corresponding phase that might be observable, provided that the sample is large enough to detect the relatively small fluctuation.

As the stars climb the RGB, the main source of energy is the H-burning shell, and convection in the outer layers becomes deeper, eventually reaching the layers nuclearly processed in previous stages and generating a discontinuity in the chemical profile ({\it first dredge-up}). The RGB--bump (Sweigart \& Gross 1978; King, Da Costa \& Demarque 1985; Renzini \& Fusi Pecci 1988; FP; Chiosi, Bertelli \& Bressan 1992) is produced when evolution up the RGB pauses as the H-burning shell passes through the discontinuity left by the deepest penetration of the convective envelope.

An {\it analogous} situation occurs at the beginning of the AGB phase: when the central abundance of He goes to zero, the He-exhausted core contracts and heats up and the H-rich envelope expands and cools so effectively that the H-burning shell extinguishes, causing the base of the convective envelope to penetrate inward again. Eventually, the expansion of the envelope is halted by its own cooling and it recontracts, the luminosity decreases and the matter at the base of the convective envelope heats up. When the H-burning shell reignites, the envelope convection moves outward in mass ahead of the H-burning shell, and the luminosity increases again. The ``slowing down'' that goes with this decreasing and posterior increasing of the luminosity must produce a second ``bump'', now in the AGB LF. Some models (e.g. Gingold 1976; Hirshfeld 1980; Bono et al. 1997) show that this phase may be accompanied by an excursion of the star to the blue during the ascent of the AGB (the so--called ``blueward noses'')\footnote{This ``blueward noses'' may eventually cross the instability strip, and this is a common way in which anomalous Cepheids are theoretically accounted for.}. 

 The behavior of the RGB and AGB LFs of a population of given age and metallicity, as predicted by the Padova isochrones (Bertelli et al. 1994, and references therein), is shown in Figure~\ref{fl004}, where the integral LFs are plotted for a population of log (age/yr)= 9.4, Z=0.004 and a Salpeter initial mass function.  The RC has $M_V \simeq$ 0. The RGB--bump appears at $M_V \simeq$ 0.5 and the AGB--bump at $M_V \simeq -0.8$, about 1 magnitude above the RC. The precise position of both features depends on metallicity and possibly age.

\section{Comparison of theoretical LFs with observed LFs: the LMC and M31} \label{compara}

To compare the characteristics of the observed AGB--bump with the characteristics expected from stellar evolution, I have computed synthetic CMDs for reasonable SFH for each system, using a recent version of the synthetic CMD simulator by Bertelli (1997, unpublished). This program interpolates within stellar evolutionary tracks of fixed mass and metallicity to determine the precise location in the CMD of stars of any mass, age and metallicity. This results in a {\it smooth} distribution of stars following a given star formation rate (SFR), initial mass function and chemical enrichment laws (see Gallart et al. 1996 for details). Also discussed are the characteristics of the AGB--bump as observed in globular clusters.   

\subsection{The LMC: the AGB--bump of an intermediate-age population} \label{compalmc}

\begin{figure}
\begin{center}
\mbox{\epsfig{file=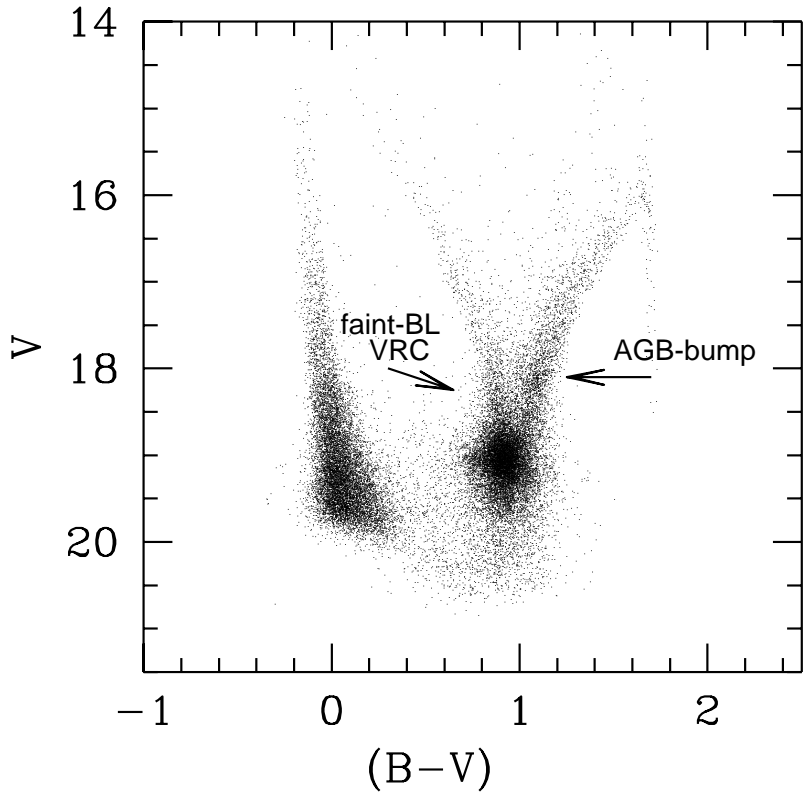,width=14cm,bbllx=2.0truecm,bblly=8.0truecm,bburx=19.0truecm,bbury=19.0truecm}}
\end{center}
\caption[]{$[(B-V), V]$ synthetic CMD reproducing the stellar populations of the LMC, computed assuming constant SFR from 3 Gyr to the present time and metallicity Z=0.001-0.006. A realistic distribution of observational effects has been assumed and simulated, using the method described in Gallart et al. (1996). A distance modulus of $(m-M)_0=18.5$ and a reddening of E(B-V)=0.1 have been assumed. The position of the AGB--bump is indicated by an arrow.}
\label{sindenis}
\end{figure}

It is generally accepted that the LMC underwent a major episode of star formation at an intermediate~age, a few Gyr ago (see Vallenari et al. 1996; Olszewski, Suntzeff \& Mateo 1996, and references therein). This is also the conclusion reached by recent {\it HST} observations of the bar and outer disk of the LMC (Elson, Gilmore \& Santiago 1997; Geha et al. 1997). Considering the information on the LMC SFH from the above references, the following SFH will be adopted to calculate the synthetic CMD: constant SFR from 3 Gyr ago to the present time and metallicity Z=0.001-0.006. The $[(B-V), M_V]$ synthetic CMD is shown in Figure~\ref{sindenis}. A realistic distribution of observational effects has been assumed and simulated, using the method described in Gallart et al. (1996). The position of the AGB--bump is indicated by an arrow. This model CMD can be compared with the Hess diagram of the LMC by Zaritsky et al. (1997, Fig.~8) and the CMD by Stappers et al. (1996, Fig.~1, but different color). The position and characteristics of the main features (main sequence, RC, tip of the RGB) are well reproduced by our model. The base of the sequence above the RC and towards the blue, produced by the BLs of stars a few Myr old, is consistent with the characteristics of the VRC noted by ZL.

\begin{figure}
\begin{center}
\mbox{\epsfig{file=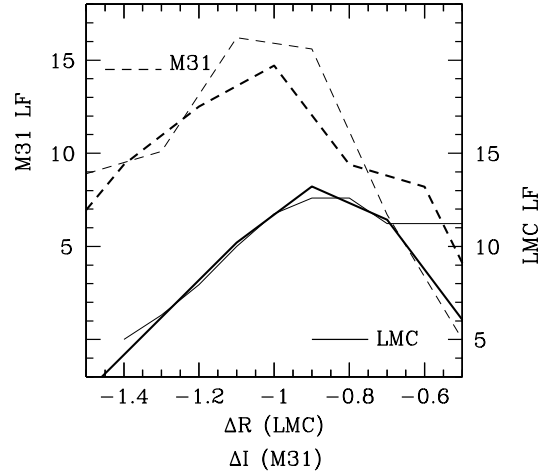,width=8cm,bbllx=2.0truecm,bblly=7.0truecm,bburx=19.0truecm,bbury=19.0truecm}}
\end{center}
\caption[]{Observed (thin lines) and synthetic (thick lines) LF for the AGB--bump area of the LMC CMD (solid lines) and the M~31 halo field CMD (dashed lines). The LFs have been normalized so that the peak number of stars in the RC are the same in both the observed and the synthetic CMD. After this normalization, the values in the $y$-axis have been set arbitrarily. The magnitudes in the $x$-axis give the position of the AGB--bump relative to the position of the RC.}
\label{lfs}
\end{figure}

The large sample of stars in the LMC CMD obtained by Alcock et al. (1997) allows the AGB--bump to be unambiguously detected, superimposed upon the RGB around $(V-R)$=0.55 mag and $R$=17.7 mag, 0.9 mag above the RC and 0.1 mag redder than it is (see their Fig. 4\footnote{This Fig. 4 appears in the version of the paper in the astro-ph archive (astro-ph/9707310), but has been removed of the ApJ version of the paper}). For both the observed and the synthetic CMDs, a LF in an interval of width $(V-R)=0.04$ mag. centered in the peak in color of the AGB--bump, has been computed. The LFs are shown in Figure~\ref{lfs}. They have been normalized so that the peak number of stars in the RC are the same for both observed and synthetic CMDs. The agreement on the position and the peak values of the AGB--bump in the LFs is remarkable.

\begin{figure}
\begin{center}
\mbox{\epsfig{file=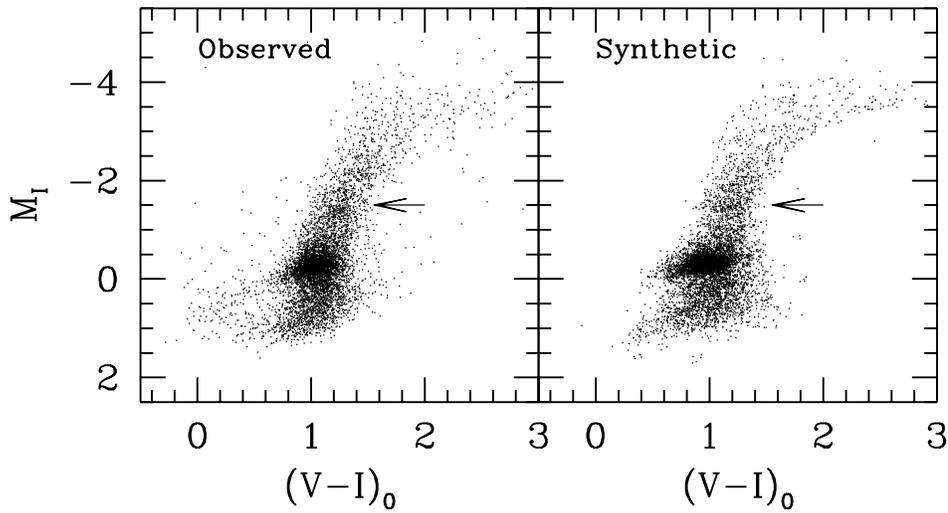,width=15cm,bbllx=0.0truecm,bblly=9.0truecm,bburx=21.0truecm,bbury=16.0truecm}}
\end{center}
\caption[]{M~31 halo field CMD and the corresponding synthetic CMD. Observational effects have been simulated in this synthetic CMD using the method described by Gallart et al. (1996) and a crowding test table obtained from a similar M31 halo field.}
\label{cmds}
\end{figure}

\subsection{The M31 Halo: the AGB--bump of an old population} \label{compam31}

The AGB--bump has also been observed for the old population in M31. This will be shown using the {\it HST} observations of the M31 halo field around the cluster G302 published by Holland et al. (1996). For this field, they find a metallicity spread of $-2.0 \le [m/H] \le -0.2$ ($0.0002 \le Z \le 0.01$). To compute the corresponding synthetic CMD, this metallicity spread has been assumed, plus a constant SFR from 8 to 10 Gyr. Figure~\ref{cmds} shows the G302 field CMD and the synthetic CMD, which includes an approximate simulation of observational effects. Apart from the differences between the two CMDs (a detailed modeling has not been attempted), the AGB--bump observed is about 1 mag. above the RC in both CMDs. 

Figure~\ref{lfs} shows the LF for the observed and synthetic population. An interval in color $\Delta(V-I)=0.4$ mag centered in the position of the AGB--bump has been used to include stars in the LF. The position and strength  of both LFs are remarkably similar. 

Note that in this case there is no VRC, as would correspond to a population with very little -if any- young population with age $\le$ 1 Gyr (see Figure~2 C and D, in Aparicio et al. 1996). 

\section{The AGB--bump in globular clusters}\label{glob}

The RGB--bump is generally observed in well--populated CMDs of globular clusters (King et al. 1985; Auri\`ere \& Ortolani 1988; FP; Ferraro et al. 1997).  The AGB--bump, however, involves a smaller fraction of stars than the RGB--bump, and generally, it is necessary to have a relatively large number of stars in the CMD to be clearly detected. In spite of that, the 66 stars in the CMD by Sandage (1970) was enough for him to include this feature in the fiducial sequences of M3 (the {\it asymptotic- branch C}). The ground-based observations of M3 (Ferraro et al. 1997) show quite clearly the AGB--bump as it is shown by most new {\it HST} CMD of the central regions of globular clusters (e.g. Rich et al. 1997).  The AGB--bump appears at the beginning of the AGB phase, as a ``clump'' of stars bluer than the RGB--AGB locus, and about $\Delta V \simeq$  1 mag. above the RC. In some cases, the substantial extension to the blue of the stars composing the AGB--bump can be related with the ``blueward noses'' (Gingold 1976; see also Section~\ref{predice}). These stars can also be related with the UV--bright stars of the Group II as defined by Zinn, Newell \& Gibson (1972).

\section{Conclusions}\label{conclu}

The AGB--bump, a relatively {\it fast} stellar evolutionary phase predicted by stellar evolutionary theory, has been identified in the CMD of a number of objects, and its presence on the CMD of the LMC has been discussed by several authors (ZL; Alcock et al. 1997; BS). Observationally, this feature can be distinguished from the well known RGB--bump because, except in very metal poor populations, the RGB--bump is observed {\it below} the RC, while the AGB--bump is always situated above it. In addition, their characteristics are clearly different. 

Using synthetic CMDs, it has been shown that the predictions by the Padova stellar evolutionary models about the position and strength of the AGB--bump are in excellent agreement with the observed structures in the LMC and in M31. This shows the accuracy with which these models are able to predict even subtle details, despite the fact that AGB evolution is still not fully understood. In the case of the LMC, the synthetic CMD also reveals a feature above and to the blue of the RC, produced by the faint BLs, with the characteristics of the VRC discussed by ZL. Thus, synthetic CMDs have proven to be a very powerful tool, not only to interpret the stellar populations of an observed CMD, but also to identify the nature of otherwise enigmatic features.  

\acknowledgments

I thank S. Holland and D. Zaritsky for sending me some of their data on M31 and the LMC. I am indebted with C. Chiosi and G. Bertelli for very important suggestions and for allowing me to use the synthetic CMDs code for this paper. It is a pleasure to thank A. Aparicio for our never--ending discussions about this paper, from the beginning to its completion, and for using his artistic abilities in Figure~1. Thanks to W. Freedman and A. Sandage for many useful discussions. In addition to A.A. and  W.F., I thank J. Dalcanton, G. Preston, I. Thompson and M. Zoccali for a critical reading of the manuscript. I would like to dedicate this paper to the memory of John Denver, whose passing saddened the completion of this work.


\begin{references}
\reference Alcock et al. 1997, \apj, 490, L59
\reference Aparicio, A., Gallart, C., Chiosi, C. \& Bertelli, G. 1996, \apj, 469, L97
\reference Auri\`ere, M. \& Ortolani, S. 1988, \aap, 204, 106
\reference Beaulieu, J. P. \& Sackett, P. D. 1997, preprint (BS)
\reference Bertelli, G., Bressan, A., Chiosi, C., Fagotto, F. \& Nasi, E. 1994, \aaps, 106, 275
\reference Bono, G., Caputo, F., Santolamazza, P., Cassisi, S. \& Piersimoni, A. 1997, \aj, 113, 2209
\reference Chiosi, C., Bertelli \& G, Bressan, A. 1992, \araa, 30, 235
\reference Da Costa, G. S., Armandroff, T. E., Caldwell, N. \& Seitzer, P. 1996, \aj, 112, 2576
\reference Dohm-Palmer et al. 1997, AJ, 114, 2527
\reference Elson, R. A. W., Gilmore, G. F. \& Santiago, B. X. 1997, \mnras, 289, 157
\reference Ferraro, F. R., Carreta, E., Corsi, C. E., Fusi Pecci, F., Cacciari, C., Buonanno, R., Paltrinieri, B., Hamilton, D. 1997, \aap, 320, 757
\reference Fusi Pecci, F., Ferraro, F. R., Crocker, D. A., Rood, R. T. \& Buonanno, R. 1990, \aap, 238, 95 (FP)
\reference Gallart, C. Aparicio, A., Bertelli, G. \& Chiosi, C. 1996, \aj, 112, 1950
\reference Geha, M. et al. 1997, AJ in press
\reference Gingold, R. A. 1976, \apj, 204, 116
\reference Hirshfeld, A. W. 1980, \apj, 241, 111
\reference Holland, S., Fahlman, G. G. \& Richer, H. B. 1996, \aj, 112, 1035
\reference King, C. R., Da Costa, G. S. \& Demarque, P. 1985, \apj, 299, 674
\reference Olszewski, E. W., Suntzeff, N. B. \& Mateo, M. 1996, \araa, 34, 511
\reference Renzini, A. \& Fusi Pecci, F. 1988, \araa, 26, 199
\reference Rich, R. M. et al. 1997, \apjl, 484, 25
\reference Sandage, A. 1970, \apj, 162, 841
\reference Stappers, B. W. et al. 1997, \pasp, 109, 292
\reference Stetson P. B. 1997, {\it Baltic Astronomy}, 6
\reference Sweigart, A. V. \& Gross P. G. 1978, \apjs, 36, 405
\reference Vallenari, A., Chiosi, C. Bertelli, G., Aparicio, A. \& Ortolani, S. 1996, \aap, 309, 367
\reference Zaritsky, D., Harris, J. \& Thompson, I. 1997a, \aj, 114, 1002
\reference Zaritsky, D., \& Lin, D.N.C. 1997b, \aj, 114, 2545 (ZL)
\reference Zinn, R. J., Newell, E. B. \& Gibson, J. B. 1972, \aap, 18, 390.
\end{references}
\end{document}